\documentclass[fleqn,usenatbib]{mnras}

\usepackage{newtxtext,newtxmath}
\usepackage[T1]{fontenc}

\DeclareRobustCommand{\VAN}[3]{#2}
\let\VANthebibliography\thebibliography
\def\thebibliography{\DeclareRobustCommand{\VAN}[3]{##3}\VANthebibliography}

\newcommand{\Msun}{\,{\rm M_\odot}}
\newcommand{\fedd}{\,{f_{\rm Edd}}}
\newcommand{\Mblack}{M_\bullet}
\newcommand{\angstrom}{\text{\normalfont\AA}}

\usepackage{graphicx}	% Including figure files
\usepackage{amsmath}	% Advanced maths commands
\usepackage{amssymb}	% Extra maths symbols
\usepackage{pgfplots}
\usepackage{subfig}
\usepackage[justification=centering]{caption}
\usepackage{pbox}

\usepackage{tipa}
\pgfplotsset{width=6.6cm,compat=1.7}  
\title[A Search for Wandering IMBHs in the Milky Way]{Detectability of Wandering Intermediate-Mass Black Holes in the Milky Way Galaxy from Radio to X-rays}

\author[B. S. Seepaul et al.]{
Bryan S. Seepaul$^{1}$\thanks{bryanseepaul@college.harvard.edu},
Fabio Pacucci$^{1,2}$\thanks{fabio.pacucci@cfa.harvard.edu} \&
Ramesh Narayan$^{1,2}$
\\
$^{1}$Center for Astrophysics $\vert$ Harvard \& Smithsonian, Cambridge, MA 02138, USA\\
$^{2}$Black Hole Initiative, Harvard University,
Cambridge, MA 02138, USA\\
}

\date{\today}

\pubyear{2022}

\begin{document}
\label{firstpage}
\pagerange{\pageref{firstpage}--\pageref{lastpage}}
\maketitle

\begin{abstract}
Intermediate-mass black holes (IMBHs, $10^{3-6} \Msun$), are typically found at the center of dwarf galaxies and might be wandering, thus far undetected, in the Milky Way (MW). We use model spectra for advection-dominated accretion flows to compute the typical fluxes, in a range of frequencies spanning from radio to X-rays, emitted by a putative population of $10^5 \Msun$ IMBHs wandering in five realistic, volume-weighted, MW environments. We predict that $\sim 27\%$ of the wandering IMBHs can be detected in the X-ray with Chandra, $\sim 37\%$ in the near-infrared with the Roman Space Telescope, $\sim 49\%$ in the sub-mm with CMB-S4 and $\sim 57\%$ in the radio with ngVLA. We find that the brightest fluxes are emitted by IMBHs passing through molecular clouds or cold neutral medium, where they are always detectable.
We propose criteria to facilitate the selection of candidates in multi-wavelength surveys. Specifically, we compute the X-ray to optical ratio ($\alpha_{\rm ox}$) and the optical to sub-mm ratio, as a function of the accretion rate of the IMBH. We show that at low rates the sub-mm emission of IMBHs is significantly higher than the optical, UV and X-ray emission. Finally, we place upper limits on the number $N_\bullet$ of these objects in the MW: $N_\bullet<2000$ and $N_\bullet<100$, based on our detectability expectations and current lack of detections in molecular clouds and cold neutral medium, respectively.
These predictions will guide future searches of IMBHs in the MW, which will be instrumental to understanding their demographics and evolution. 
\end{abstract}


%%%%%%%%%%%%%%%%%%%%%%%%%%%%%%%%%%%%%%%%%%%%%%%%%%

%%%%%%%%%%%%%%%%% BODY OF PAPER %%%%%%%%%%%%%%%%%%

\section{Introduction}
\label{sec:intro}
Black holes (BHs) are arguably paradoxical objects: while light cannot escape from their event horizons, the regions of space around them can be the brightest environments in the Universe. Currently we have detected quasars, or super-massive BHs with mass $\Mblack > 10^6 \Msun$, as bright as $\sim 10^{48} \, \mathrm{erg \, s^{-1}}$ \citep{Wu_2015}. This power is generated by an influx of material falling in the gravitational field of the massive object, and forming an accretion disk, which ultimately transforms kinetic energy into radiation. In the standard $\alpha$-disk model \citep{Shakura_Sunyaev_1973, Novikov_Thorne}, typically $\sim 10\%$ of the rest-mass energy of the infalling matter, $Mc^2$, is transformed into radiation. Labeling this matter-to-energy conversion efficiency as $\epsilon$ and the mass accretion rate as $\dot{M}$, the emitted power is then $L = \epsilon \dot{M} c^2$ (see, e.g., the review on accretion by \citealt{Narayan_2005}).

The value $\epsilon \sim 0.1$ provides an effective description of the matter-to-energy efficiency for an accretion process that occurs with rates comparable to the Eddington rate $\dot{M}_{\rm Edd} = 1.4\times 10^{18} \Mblack \, \mathrm{g \, s^{-1}}$, and $\Mblack$ is expressed in solar masses, typically from $\dot{M} \sim 0.01 \dot{M}_{\rm Edd}$ to $\dot{M} \sim \dot{M}_{\rm Edd}$. For super-Eddington ($\dot{M} \gg \dot{M}_{\rm Edd}$) rates \citep{Volonteri_2005}, the structure of the accretion disk is modified due to advection, which ultimately decreases the amount of energy radiated away. Several studies of the so-called slim-disk solution have shown that the radiative efficiency $\epsilon$ can be significantly lower than $10\%$ (e.g., \citealt{Begelman_1978, Paczynski_1982, Abramowicz_1988, Sadowski_2009}). In addition to analytical slim disk models, we now also have general-relativistic magneto-hydrodynamic (GRMHD) simulations of super-Eddington accretion with radiation, which show that the flow is radiatively inefficient \citep{Sadowski_2014}. 

On the other side of the accretion spectrum, i.e. for $\dot{M} \ll \dot{M}_{\rm Edd}$, radiative efficiencies also decrease dramatically as we enter the domain of advection-dominated accretion flows, or ADAF \citep{Narayan_1994, Narayan_1995, Abramowicz_1995, Narayan_2008, Yuan_Narayan_2014}. BHs accreting in ADAF mode have radiative efficiencies possibly several orders of magnitude lower than the standard $\sim 10\%$ value. As the conditions for large accretion rates are rare, especially at low redshift (e.g., \citealt{Power_2010}), most of the BHs in the Universe accrete in ADAF mode, including the super-massive BH at the center of our Galaxy (e.g., \citealt{Yuan_2003}).
Due to mass loss and convection typical in ADAFs, complex numerical simulations are required to study this accretion mode. From the pioneering, two-dimensional simulations of \cite{Stone_1999}, the field has advanced significantly, with recent GRMHD simulations able to probe accretion rates of $\dot{M} \sim 10^{-9} \dot{M}_{\rm Edd}$ and radiative efficiencies $\epsilon \sim 10^{-6}$ \citep{Ryan_2017, Ressler_2017, Chael_2018, Chael_2019}.

In addition to the majority of stellar-mass and super-massive BHs, ADAF models are also suitable to describe the accretion flows onto putative intermediate-mass BHs (IMBHs) wandering in the Milky Way (MW) Galaxy. IMBHs bridge the gap between stellar mass and super-massive objects, and are typically in the mass range $10^3 \Msun < \Mblack <  10^6 \Msun$, but the extremes can vary depending on the definition. IMBHs have been detected extensively in dwarf galaxies (see, e.g., the recent review \citealt{Greene_2020_IMBH}), which have active fractions typically in the range $\sim 5\%-22\%$ \citep{Pacucci_2021_active}.
Note that this is in accordance with standard scaling relations (see, e.g., \citealt{Kormendy_Ho_2013}). 

Recent studies (e.g., \citealt{Bellovary_2010, Gonzales_2018, Greene_2021_wandering, Ricarte_2021b, Ricarte_2021a}) suggest that IMBHs may be wandering in the MW. These objects could have originated via several mechanisms, also in a dwarf that is subsequently captured by the MW. Examples of formation channels are: (i) super-Eddington accretion onto stellar-mass BHs (e.g., \citealt{Ryu_2016}), (ii) direct collapse of high-mass quasi-stars (e.g., \citealt{Volonteri_2010_IMBH, Schleicher_2013}, (iii) runaway mergers in dense globular stellar clusters (e.g., \citealt{PZ_2002, Gurkan_2004, Shi_2021, Gonzalez_2021}), and, in the high-$z$ Universe, (iv) supra-exponential accretion on seed black holes (e.g., \citealt{Alexander_Natarajan_2014, Natarajan_2021}).

However they are formed, these wandering IMBHs would be accreting from the inter-stellar medium (ISM) at low-rates, thus in ADAF mode. Their accretion rate can be modeled by a Bondi rate \citep{Bondi_1952}, rescaled to take into account outflows and convection, as shown by \cite{Igumenshchev_2003} and \cite{Proga_2003}, and used for the neutron star case by \cite{Perna_2003}. The observability of wandering massive BHs in nearby galaxies and globular clusters, especially using radio techniques, is the topic of recent attention, which highlights the importance of this topic (see, e.g., \citealt{Nyland_2018, Plotkin_2019, Guo_2020, Wrobel_2021}).

In this study, we investigate the radiative properties, from the X-ray to the radio bands, of putative IMBHs of $\sim 10^5 \Msun$ wandering in the MW, in five of its typical ISM environments. The goal is to inform and guide future observational efforts to detect these sources and bridge the gap in the BH population of our Galaxy.

\section{Methodology}
\label{sec:methods}
In this Section we discuss the methodology used in this study. In particular, we describe the accretion model, the five physical environments of reference for the MW, the methods to compute the spectral energy distribution (SED), the velocity distribution of the population of IMBHs, and their spatial distribution. Finally, we describe the detectors considered in this study.

\subsection{Accretion Model}
The accretion rate of an object spherically accreting from its environment can be modeled by the Bondi accretion model \citep{Bondi_1944, Bondi_1952}:
\begin{equation}
    \dot{M}_{\rm B} = \frac{4 \pi G^{2} \Mblack^{2} \rho}{(v^{2} + c_{s}^{2})^{3/2}} \, ,
\end{equation}
where $\rho$ and $c_{s}$ are the density and sound speed of the ambient ISM, $v$ is the relative speed of the BH with respect to the medium, and $\Mblack$ is the BH mass. For the remainder of this study, we keep $\Mblack$ constant at $10^{5} \, \Msun$, as a typical mass for IMBHs \citep{Greene_2020_IMBH}. Also, note that in the original formulation by Bondi, the accretion is spherically symmetric and the accretor is at rest with respect to the gas. Here, we use the Bondi-Hoyle-Lyttleton model which includes a velocity $v$ with respect to the medium (see, e.g., the review by \citealt{Bondi_review_2004} on the subject).

As discussed in \S \ref{sec:intro}, over the years several corrections to the standard Bondi model have been implemented, to take into account effects that limit the rate effectively falling onto the BH \citep{Igumenshchev_2003, Proga_2003}. First, the presence of turbulent outflows: as ambient matter is pulled closer to the event horizon, a substantial amount could be lost as an outflow to the external environment \citep{Blandford_Begelman_1999}. Second, the convection within the accretion disk around the BH, which provides a new channel for the outward transport of energy that in turn changes the nature of the solution and the accretion rate \citep{Narayan_2000, Quataert_Gruzinov_2000}.

To account for these effects, we adopted the following correction to the standard Bondi accretion rate \citep{Igumenshchev_2003, Proga_2003}:
\begin{equation}
    \dot{M} = \Big(\frac{R_{\rm in}}{R_{\rm a}}\Big)^{p} \dot{M}_{\rm B} \, .
    \label{eq:Bondi_corr}
\end{equation}
Here, we adopt a value $R_{\rm in} = 50 \times R_{\rm S}$ \citep{Abramowicz_2002} for the effective inner radius, where $R_{\rm S}$ is the Schwarzschild radius of the IMBH:
\begin{equation}
    R_{\rm S}  = \frac{2 G \Mblack}{c^{2}}.
\end{equation}
$R_{\rm a}$ is the accretion radius, defined as the distance from the central mass where the ambient matter feels an inward gravitational force from the BH (see \citealt{Bondi_1952} and further corrections for the Bondi-Hoyle-Lyttleton model in \citealt{Bondi_review_2004}):
\begin{equation}
    R_{\rm a} = \frac{2 G \Mblack}{v^2 + c_{s}^{2}} \, .
\end{equation}
The index $p$ determines how the accretion flow transitions to sub-Bondi rates for radii smaller than the accretion radius.
We set the index $p=0.5$, following simulations that suggest that $p\sim 0.5 - 1$ \citep{Pen_2003, Yuan_2012, Yuan_Narayan_2014}. Specifically, the recent study by \cite{Ressler_2020} uses a value of $p=0.5$, while $p\sim 0.3$ applies to Sgr A* \citep{Yuan_2003}.

As $c >> c_{s}$, $R_{\rm in} << R_{\rm a}$, hence $\dot{M}_{\rm B}$ is scaled down to $\dot{M}$. Notably, at radii $r > R_{\rm a}$, the accretion rate is similar to the Bondi rate, while at $r < R_{\rm a}$ the accretion rate is reduced due to the presence of turbulent outflows and convection.

Note that our choice of $p=0.5$, while consistent with numerical simulations and scanty observations \citep{Pen_2003, Yuan_2012, Yuan_Narayan_2014, Ressler_2020}, is somewhat arbitrary. A value of $p > 0.5$ would decrease the accretion rate on the IMBH, consistently decreasing all predicted fluxes. In \S \ref{sec:results} we show how a different choice, namely $p=0.75$, would affect our predictions for the accretion rates on our putative population of IMBHs.

\subsection{ISM Environments in the MW}
Using the modified version of the Bondi accretion rate expressed in Eq. \ref{eq:Bondi_corr}, we note that its value depends on three parameters: (i) the density of the ambient medium $\rho$, (ii) the speed of sound within the ambient medium $c_{s}$, and (iii) the velocity of the IMBH with respect to the surrounding ISM $v$. The density and the sound speed are both properties of the ambient medium. Thus, to obtain specific values for $\rho$ and $c_{s}$, we focused on five baseline environments of the ISM of the MW: molecular cloud (MC), hot ionized medium (HIM), warm neutral medium (WNM), warm ionized medium (WIM), and cold neutral medium (CNM). Table \ref{table:summary_ISM} shows typical parameters that characterize each of the five environments studied (see, e.g., \citealt{Ferriere_2001}).

\begin{table*}
\begin{tabular}{ |p{2.0cm}||p{3.5cm}|p{4.5cm}|p{2.5cm}|  }
 \hline
\textbf{Environment}& \textbf{Composition} & \textbf{Number Density} [$\mathrm{cm^{-3}}$] & \textbf{Temperature} [K]\\
 \hline
 \textbf{MC} & $H_{2}$ & $10^{2}$ -- $10^{4}$  cm$^{-3}$ & 10 -- 20 K\\
 \textbf{CNM} & HI & 20 -- 50 cm$^{-3}$ & 80 -- 100 K\\
 \textbf{WNM} & HI & $3 \times 10^{-1}$ -- $5 \times 10^{-1}$ cm$^{-3}$ &  $6 \times 10^{3}$ -- $1 \times 10^{4}$ K\\
 \textbf{WIM} & $H^{+}$ & $1 \times 10^{-1}$ -- $5 \times 10^{-1}$ cm$^{-3}$& $6 \times 10^{3}$ -- $1.2 \times 10^{4}$ K\\
 \textbf{HIM} & $H^{+}$ & $2.5 \times 10^{-3}$ -- $3.5 \times 10^{-3}$ cm$^{-3}$ & $10^{6}$ -- $10^{7}$ K\\
 \hline
\end{tabular}
\caption{\label{tab: Table 1} Summary of the typical physical properties of the five ISM environments studied. The columns display the name of the environment, its main molecular composition, as well as their typical number density and temperature ranges.}
\label{table:summary_ISM}
\end{table*}

The sound speed $c_{s}$ is derived as:
\begin{equation}
    c_{s} = \sqrt{\frac{\gamma k_{\rm B} T}{\mu m_{H}}},
    \label{eq:sound_speed}
\end{equation}
where $\gamma$ is an adiabatic gas constant dependent upon the specific heat of the gas under constant temperature and pressure. For a diatomic gas ($H_{2}$) $\gamma = 7/5$, while for a monoatomic gas such as HI or $H^{+}$ $\gamma = 5/3$. Finally $T$ is the temperature of the medium while $k_{\rm B}$ is the Boltzmann constant. For each environment, we implemented a Gaussian distribution of temperatures, with the $\pm 3\sigma$ values corresponding to the minimum and maximum of the ranges displayed in Table \ref{table:summary_ISM}. $\mu m_{H}$ is the mass of the molecule that each medium is composed of, expressed in terms of the mass of a hydrogen atom. For the molecular cloud, which is mostly comprised of $H_{2}$ molecules, $\mu = 2$; for the CNM and WNM, which are mostly comprised of $HI$ molecules, $\mu = 1$; and for the WIM and HIM, which are mostly comprised of $H^{+}$ molecules, $\mu = 1/2$. Substituting these parameters into Eq. \ref{eq:sound_speed}, we obtain Gaussian distributions for the sound speed $c_{s}$ in each environment. Note that lower temperatures are associated with regions where molecular hydrogen, a strong coolant, is present.

We calculated the mass density $\rho$ of each environment from its number density distribution and molecular composition, displayed in Table \ref{table:summary_ISM}. Hence, we modelled the mass density using a Gaussian distribution between the minimum and maximum values of $\rho$ (calculated from the minimum and maximum of $n$). If $n$ is the number density distribution, in a specific environment with mean density $\bar{n}$, the Gaussian distribution centered around $\bar{n}$ is set as to have $\max(n)$ and $\min(n)$ as the $+3\sigma$ and $-3\sigma$ values of the distribution, respectively. In the case of MC environments, with the number density varying over several orders of magnitude, the distribution is log-normal.

Finally, we considered volume fractions of the MW occupied by each of the ISM environments studied (see Table \ref{table:volume_fractions}). These volume fractions, derived from \cite{Ferriere_2001}, indicate that, e.g., MC environments are extremely rare, with only $0.05\%$ in volume of the MW. On the contrary, the most common environment is the HIM, which occupies almost half of the MW volume. Note that these volume fractions refer to the gaseous regions of the MW, i.e., the disk and the central volume. In the calculation of these fractions, the volume pertaining to the MW halo is not considered \citep{Ferriere_2001}.

Note that the WNM and WIM environments, despite their different ionization states, are very similar for our purposes, with comparable number densities, temperatures and even MW volume fractions.

\begin{table}
    \centering
    \begin{tabular}{c|c}
    \hline
\textbf{Environment} & \textbf{\% Volume of MW}\\
 \hline
        \textbf{MC} &  $\sim 0.05\%$\\
        \textbf{CNM} & $\sim 1.0\%$\\
        \textbf{WNM} & $\sim 29.95\%$\\
        \textbf{WIM} & $\sim 22.0\%$\\
        \textbf{HIM} & $\sim 47.0\%$\\
    \hline
    \end{tabular}
    \caption{Summary of the MW volume fractions for each of the five ISM environments studied.}
    \label{table:volume_fractions}
\end{table}

\subsection{Velocity Distribution of IMBHs}
\label{subsec:vel}
The final step to calculate the accretion rate on wandering IMBHs involves constructing an appropriate model for the relative speed $v$ of these objects with respect to the local gas. 

An educated guess for the velocity distribution of IMBHs in the MW compares them to the population of high-velocity stars (see, e.g., the classic catalog by \citealt{Roman_1955} and a more recent study such as \citealt{Du_2018}). This old population of stars does not share the common velocity of disk stars around the center of the Galaxy, and often have very elliptical or even tilted orbits with respect to the disk. The high-velocity stars that populate the MW halo have typical galactocentric velocities of $\sim 200 \, \mathrm{km \, s^{-1}}$.

\begin{figure}
\includegraphics[width=\columnwidth]{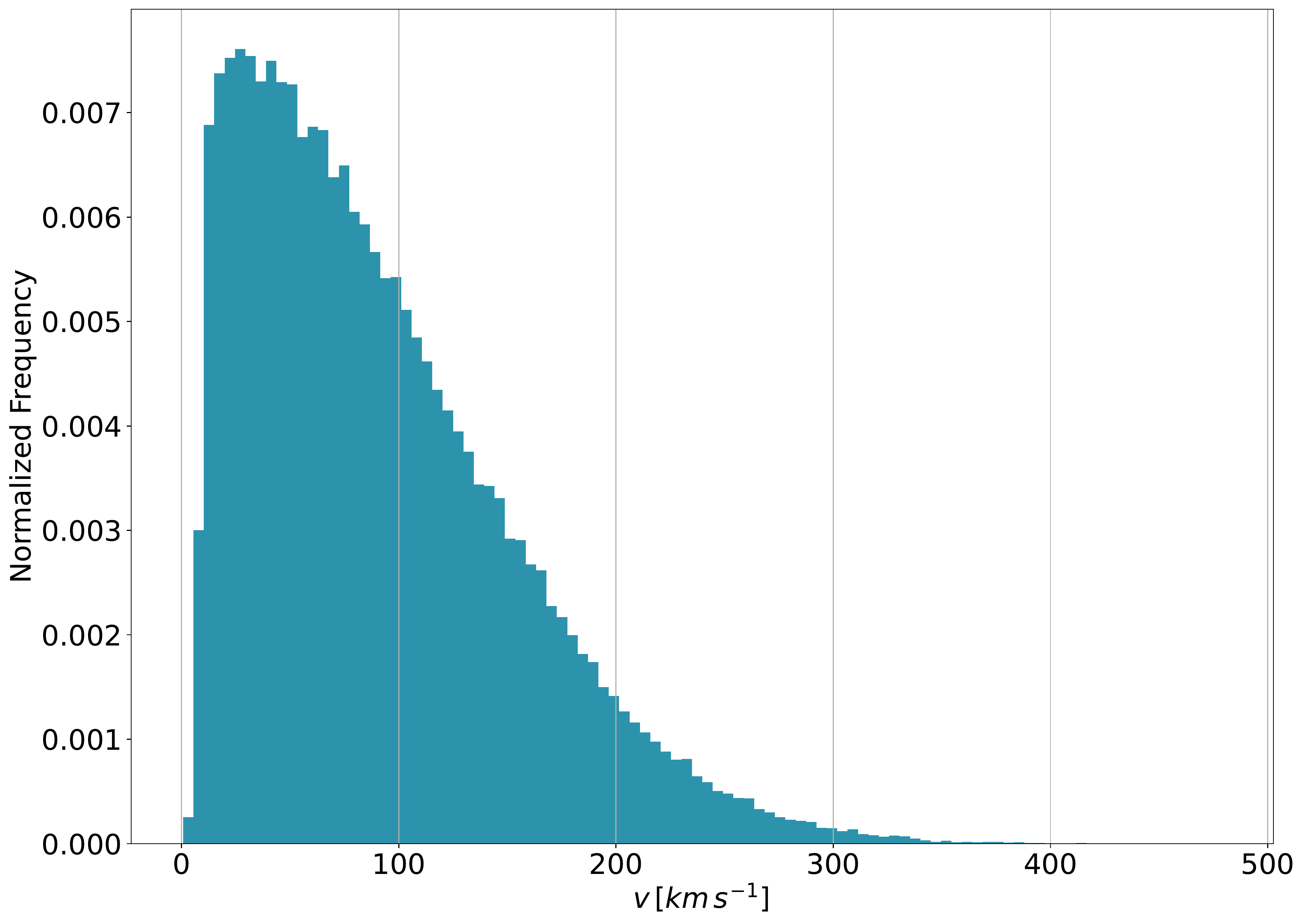}
    \caption{Velocity distribution of IMBHs relative to the surrounding gas obtained from Illustris TNG50 simulations. The mean velocity is $\sim 88\, \mathrm{km \, s^{-1}}$ and the distribution is significantly skewed towards high velocities.}
    \label{fig:v_pdf}
\end{figure}

While the velocity distribution of high-velocity stars could in principle be a good proxy for the description of the velocity of IMBHs, in our case we are mostly interested in the velocity with respect to the local gas, not in the galactocentric velocity. For this reason, we employ the probability distribution function of the velocity of IMBHs with respect to the local gas frame found in \cite{Weller_2022}. This study employed the Illustris TNG50 \citep{Pillepich_2018, Nelson_2018} suite of cosmological simulations to investigate the distribution of IMBH velocities. Remarkably, this study found that the velocity distribution relative to gas local to the IMBH has a mean of $\sim 88 \, \mathrm{km \, s^{-1}}$ and is significantly skewed towards high velocities (see Fig. \ref{fig:v_pdf}).
The velocity of the IMBH with respect to the local gas frame was calculated from the mean of the velocity vectors relative to each of the four closest gas particles in the simulation \citep{Weller_2022}.
The fact that the velocity with respect to the local gas is lower than expected boosts our predictions for the accretion rate from the ISM onto these IMBHs, as the Bondi rate scales as $\propto v^{-3}$. Additionally, \cite{Weller_2022} concluded that wandering IMBHs should preferentially be found in the central $\sim 1$ kpc. As discussed in \S \ref{sub:spatial}, MC environments are more likely to be located in the central region. As wandering IMBHs are always observable in MCs (see \S \ref{sec:results}), this boosts our chances of detecting them.

\subsection{Spectral Energy Distribution of IMBHs}
\label{subsec:SED}
\begin{figure}
\includegraphics[width=\columnwidth]{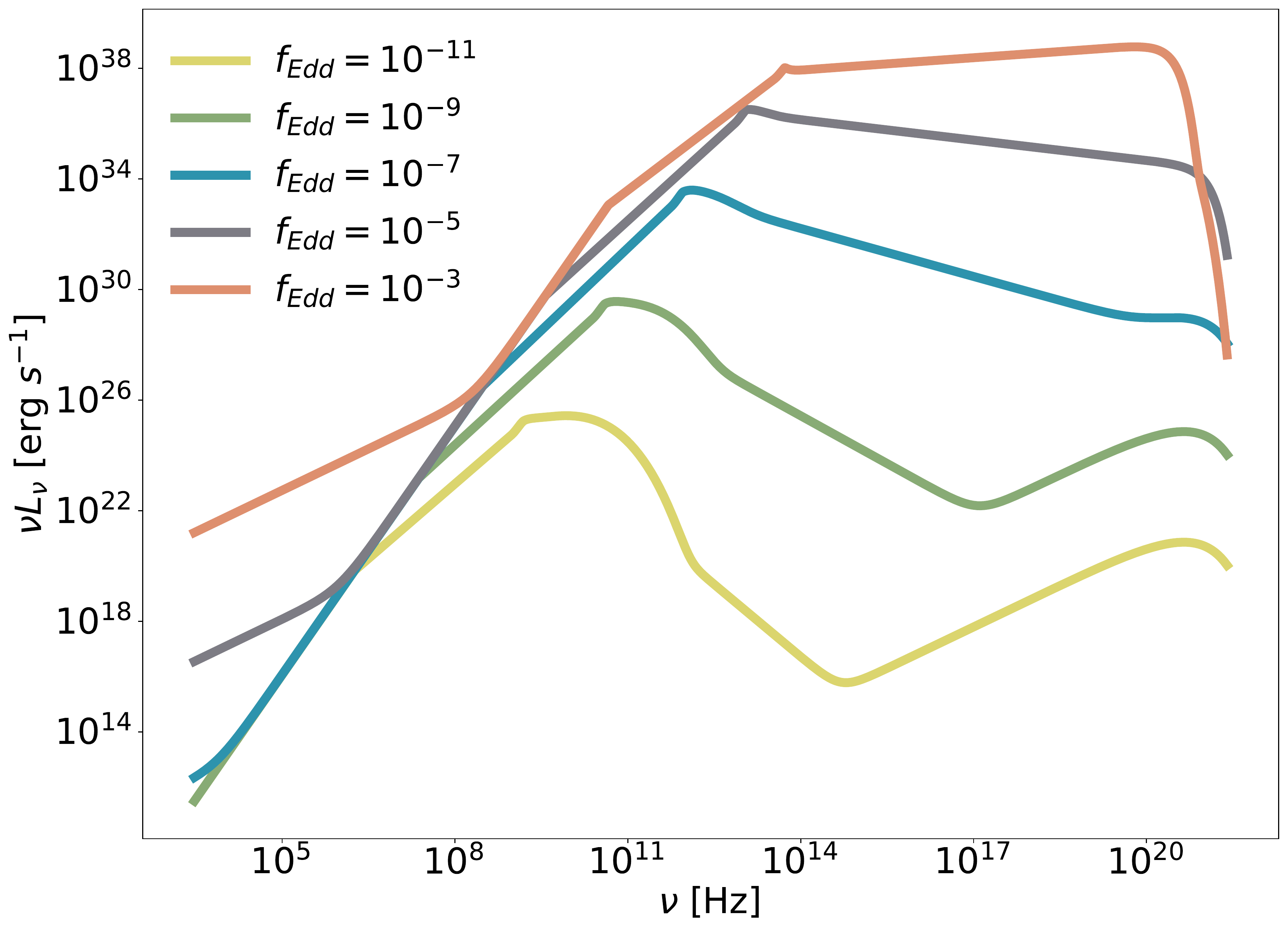}
    \caption{Display of SEDs for a $10^5 \Msun$ IMBH with five Eddington ratios in the range $10^{-11}$ to $10^{-3}$.}
    \label{fig:SED_display}
\end{figure}

The spectral energy distribution (SED) of wandering IMBHs was calculated using a code specifically designed for compact objects accreting in ADAF mode. The code and its physical prescriptions are presented in \cite{Pesce_2021}. The code is based on a physical model specifically tailored to ADAF accretion and outputs the SED for a black hole, given its mass $\Mblack$, which we fix at $10^5\Msun$, and the accretion rate $\dot{\Mblack}$, computed following the prescriptions outlined in the previous three subsections. 
The SEDs used in this work result from the combination of synchtrotron, bremsstrahlung, and inverse Compton radiation, as detailed in the appendix of \cite{Pesce_2021}. As such, the radio emission considered here derives from thermal emission associated with the accretion flow. Figure \ref{fig:SED_display} displays a collection of SEDs, with Eddington ratios (defined as the accretion rate normalized to the Eddington rate) ranging from $10^{-11}$ to $10^{-3}$.

The maximum Eddington ratio that this code can handle, given the ADAF mode assumption, is $\mathrm{Log_{10}}\fedd =-1.7$. Using higher values would yield unphysical results.
Hence, we ignore all models that predict an Eddington ratio above $\mathrm{Log_{10}}\fedd =-1.7$, a fraction $\sim 7\times 10^{-6}$ of the total in our study.

\subsection{Spatial Distribution of IMBHs}
\label{sub:spatial}
In the previous Section we described how we obtained the SED, expressed as the monochromatic luminosity $L_{\nu}$ in a given frequency range. However, to calculate the flux we need to specify the distance $d$ between the source and the observer. 

First, we considered a putative population of $10^6$ IMBHs, equally distributed in the entire volume of the MW, such that a number of IMBHs occupies each ISM regions proportionally to their volume fractions indicated in Table \ref{table:volume_fractions}. Hence, only a small number ($\sim 500$) of our $10^6$ IMBH probes will be in MCs, while $\sim 47\%$ of them will be inside the HIM region. It is important to remark that we are not suggesting that $10^6$, or any specific number of, IMBHs populate the MW. This number was chosen only to provide a representative statistical sample of fluxes for our study, while keeping the computation time to a manageable level. See \S \ref{subsec:constraints} for a discussion on some upper limits on the number of IMBHs in the MW derived from this study.

For the environments of the MW that consist of HI and HII, i.e., CNM, WNM, WIM, and HIM, we assumed that they were uniformly distributed throughout the MW Galaxy \citep{Ferriere_2001}. To implement this model, we draw random distances between $0.1$ kpc (the outer edge of the Local HI bubble) and $8$ kpc (the Galactic center), again assuming that the IMBHs are uniformly distributed in each environment. Note that with this assumption we are limiting the distance at which we can observe an IMBH up to the Galactic center. 

Regarding MC environments, recent studies (see, e.g., the review by \citealt{MC_2015}) suggest that most molecular clouds are found in a volume within $\sim 4 \, \mathrm{kpc}$ from the Galactic center, which is $\sim 8$ kpc away from the solar system. For this reason, in our code, we decided to use a mean distance $d = 8$ kpc to model the fluxes of IMBHs coming from MC environments.

With these distances accounted for, our machinery to model the IMBH flux distributions within each environment of the MW is complete.

\subsection{Description of the Detectors Considered}
Here we provide a short description of the detectors considered in this study, both currently operative and prospective.
This description is not exhaustive, and the interested reader is referred to the cited references for a much broader account of their technical features. In addition, Table \ref{table:detectors} reports the most relevant properties of each detector described here.

\begin{table*}
    \centering
    \begin{tabular}{c|c|c|c|c}
    \hline
\textbf{Detector} & \textbf{Frequency Range} [Hz] & \textbf{Wavelength Range} [$\mathrm{\mu m}$] & \textbf{Flux Limit [$\mathrm{erg \, s^{-1} \, cm^{-2}}$]} & \textbf{References}\\
\hline
\multicolumn{5}{|c|}{Currently operative detectors} \\
    
 \hline
        \textbf{Chandra} & $4.8 \times 10^{16}$ -- $2.4 \times 10^{18}$ & $1.2 \times 10^{-4}$ -- $6.2 \times 10^{-3}$ & $2.0\times10^{-16}$ & \cite{Weisskopf_2000}\\
        \textbf{eROSITA} & $4.8 \times 10^{16}$ -- $2.4 \times 10^{18}$ & $1.2 \times 10^{-4}$ -- $6.2 \times 10^{-3}$ & $2.3 \times 10^{-14}$ & \cite{Predehl_2021}\\
        \textbf{Planck (Channel 6)} & $1.8 \times 10^{11}$ -- $2.8 \times 10^{11}$ & $1.1 \times 10^{3}$ -- $1.7 \times 10^{3}$ & $1.8 \times 10^{-13}$ & \cite{Plack_HFI_Core_Team_2011} \\
        \textbf{SCUBA} & $3.5 \times 10^{11}$ -- $6.7 \times 10^{11}$ & $450$ -- $850$ & $6.5 \times 10^{-16}$ & \cite{Scuba_2013} \\
         
\hline
\multicolumn{5}{|c|}{Prospective detectors} \\
    \hline
    \textbf{AXIS} & $1.2 \times 10^{17}$ -- $4.8 \times 10^{17}$ & $6.2 \times 10^{-4}$ -- $2.5 \times 10^{-3}$ & $3.0\times10^{-18}$ & \cite{AXIS_2018}\\
    \textbf{Roman HLSS} & $1.5 \times 10^{14}$ -- $3.0 \times 10^{14}$ & $1.0$ -- $1.93$ & $1.0\times 10^{-16}$ & \cite{HLSS_2021}\\
    \textbf{CMB-S4 (Channel 11)} & $2.4 \times 10^{11}$ -- $3.0 \times 10^{11}$ & $1000$ -- $1250$ & $2.4 \times 10^{-15}$ & \cite{CMB-S4}\\
    \textbf{ngVLA} & $1.2 \times 10^{9}$ -- $1.2 \times 10^{11}$ & $2.5 \times 10^{3}$ -- $2.5 \times 10^{5}$ & $3.5 \times 10^{-19}$  & \cite{ngVLA}\\
    \hline

    \end{tabular}
    \caption{Summary of the technical features of the instruments considered in this study. From left to right: name of the detector, its range of frequencies and wavelengths, flux limits and reference paper. For the ngVLA, see also the \href{https://ngvla.nrao.edu/page/performance}{updated performance estimates}.}
    \label{table:detectors}
\end{table*}

In the X-ray band, we considered the extended ROentgen Survey with an Imaging Telescope Array (eROSITA) and the Chandra X-ray Observatory, both in the $0.2-10$ keV band. Additionally, we considered the proposed Advanced X-ray Imaging Satellite (AXIS) detector. While Chandra and AXIS are designed for targeted observations with a significantly higher sensitivity and angular resolution, eROSITA has the advantage of an all-sky survey \citep{Weisskopf_2000, AXIS_2018, Predehl_2021}.

In the near-infrared (NIR) band, we considered the Roman Space Telescope, with the planned program HLSS (high-latitude spectroscopic survey, \citealt{HLSS_2021}). This survey, in the $0.93-2.00 \, \mu$m range, offers an optimal balance between depth and a wide area of $\sim 1700 \, \mathrm{deg^2}$ \citep{HLSS_2021}.

Finally, in the sub-mm/radio bands, we considered four detectors: the Submillimetre Common-User Bolometer Array (SCUBA, \citealt{Scuba_2013}), Planck (channel 6 of the high-frequency instrument, HFI, centered at 217 GHz, \citealt{Plack_HFI_Core_Team_2011}), the Cosmic Microwave Background Stage 4 (CMB-S4, channel 11, centered at 270 GHz, of the proposed CMB experiment, see, e.g., \citealt{CMB-S4}) and the next-generation Very Large Array (ngVLA, a proposed next-generation radio telescope, see, e.g., \citealt{ngVLA}).
We use the full frequency coverage of ngVLA, between 1.2 GHz and 116 GHz.

With this choice of detectors, we provided a good balance between wide surveys with different depths (CMB-S4, Planck, SCUBA, Roman, eROSITA) and targeting instruments with a very deep flux limits, such as ngVLA, Chandra and AXIS. It is important to point out that this list of instruments used is not intended to be exhaustive. Our goal was to study a manageable number of detectors that span the widest electromagnetic (EM) spectrum possible, with very different operating modes (targeting vs. survey), and different sensitivities.

Regarding the flux limits reported in Table \ref{table:detectors}, we used values retrieved from the technical reference literature. As the technical material is diverse, standards to express the limits vary. For instance, regarding the prospective detectors, the sensitivity used for the Roman HLSS is a $6.5\sigma$ limit, while for the CMB-S4, it is a $5\sigma$ one. The interested reader is encouraged to review the technical details in the referenced literature. 

A discussion on the different angular resolutions characterizing the chosen instruments is warranted. Focusing on the prospective detectors, the angular resolutions span from the $\sim 1$ arcmin of CMB-S4, to the $\sim 0.5$ arcsec of AXIS, to the $\sim 0.1$ arcsec of Roman, reaching down to $\sim 1$ mas of the ngVLA. Of course, such a large span in angular resolutions will affect the detection prospects of wandering IMBHs or even their photometric study. Source confusion from, e.g., star-forming regions and Galactic cirrus emissions may impact the photometric measurements with lower-resolution instruments.

Additional telescopes targeting thermal emission will also play an important role in the search for Galactic IMBHs, such as ALMA \citep{ALMA}, SPHERE-X \citep{SphereX}, EUCLID \citep{Euclid}, JWST \citep{JWST}.

\section{Results}
\label{sec:results}
In this Section we discuss the detectability of IMBHs wandering in the MW Galaxy by current and future observatories in the following bands of the EM spectrum: X-ray, NIR, sub-mm and radio. In the next pages, we display the probability density distribution (PDF) for the observed fluxes generated by our distribution of IMBHs located in each of the five MW environments modeled. The fluxes of each environment are weighted by the appropriate percent-volume they occupy in the MW (see Table \ref{table:volume_fractions}).

In each figure, the dotted vertical lines represent flux limits for detectors in a specific part of the EM spectrum. We define the detectability fraction as the percentage of the population of $10^6$ sampled IMBHs whose flux in a band is above the flux limit.

\subsection{Accretion Rates Distribution}
\begin{figure}
\includegraphics[width=\columnwidth]{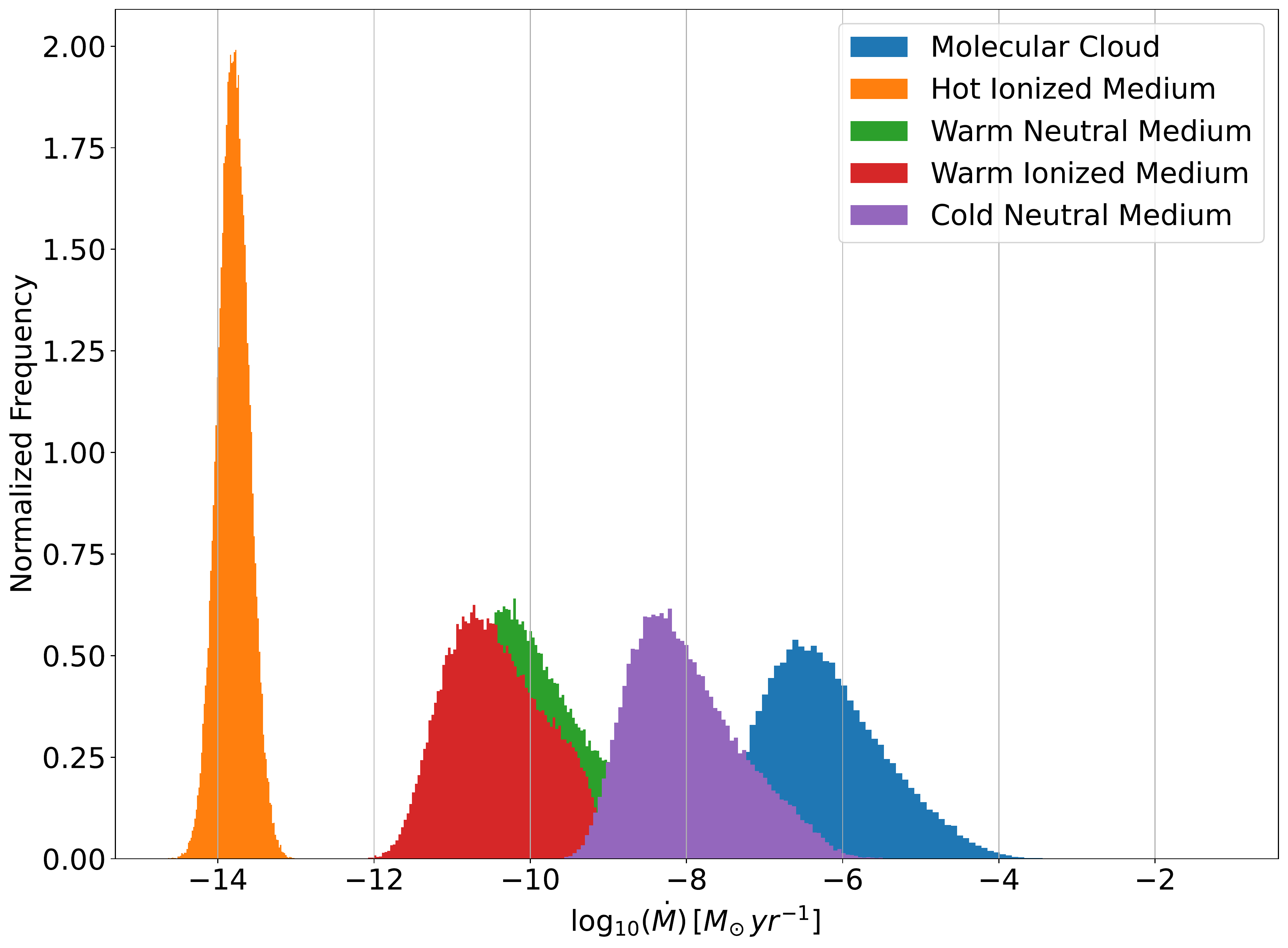}
\includegraphics[width=\columnwidth]{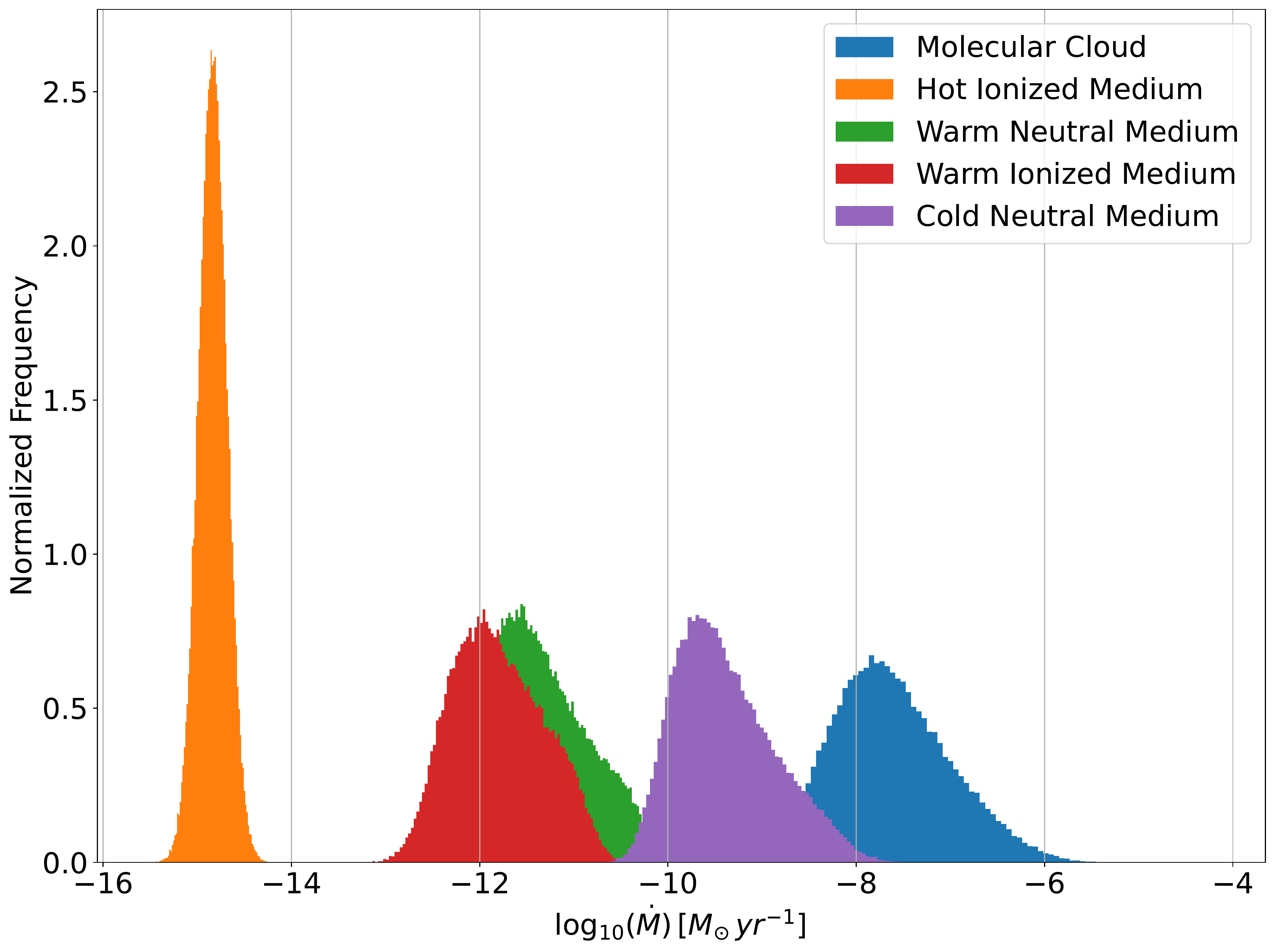}
    \caption{Distribution of accretion rates in $\mathrm{\Msun \, yr^{-1}}$ for the five environments investigated. Each distribution is normalized such that it integrates to 1. \textbf{Top}: Distribution for $p=0.5$. \textbf{Bottom}: Distribution for $p=0.75$. The overall decline of all the accretion rates predicted is evident.}
    \label{fig:rates_pdf}
\end{figure}

Before delving into observability estimates, we focus on accretion rates predictions categorized in the five MW environments investigated. The top panel of Fig. \ref{fig:rates_pdf} shows that predicted accretion rates span $\sim 10$ orders of magnitude, from $10^{-14}$ to $10^{-4} \mathrm{\Msun \, yr^{-1}}$. As expected from the density and temperature values reported in Table $\ref{tab: Table 1}$, MC and CNM environments show the largest accretion rates, followed by WNM, WIM and HIM. As we show in the following sections, MC and CNM consistently provide the best MW environments for detecting these sources, in any frequency range investigated.

Note that the Eddington rate on an IMBH of $10^5\Msun$ is $\dot{M}_{\rm Edd} \approx 2 \times 10^{-3} \mathrm{\Msun \, yr^{-1}}$. Given that we are constrained to Eddington ratios $<10^{-1.7}$ (see \S \ref{subsec:SED}), we can correctly model accretion rates up to $\dot{M} \approx 10^{-4.3} \mathrm{\Msun \, yr^{-1}}$, which means that we are discarding from our calculation the few rates above this limit (a fraction $\sim 7\times 10^{-6}$ of the total).

To investigate the effect of a change in the parameter $p$ in Eq. \ref{eq:Bondi_corr}, in the bottom panel of Fig. \ref{fig:rates_pdf} we show the equivalent of the top panel of the same figure, but calculated with $p=0.75$. The overall change shifts the accretion rates by $1-2$ orders of magnitude to lower values.

\subsection{X-ray and NIR Emission from IMBHs}

\begin{figure*}
\includegraphics[width=\columnwidth]{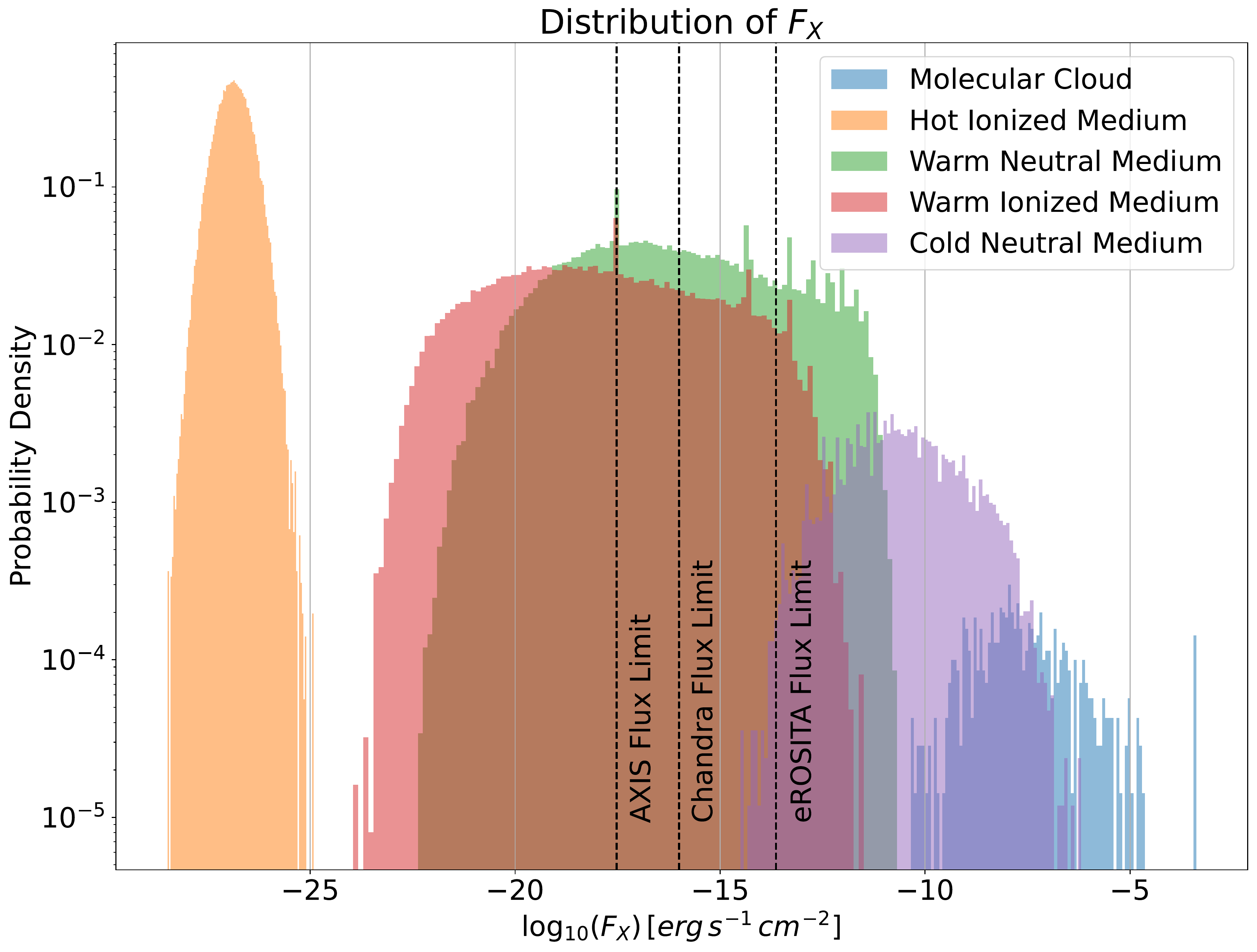}
\includegraphics[width=\columnwidth]{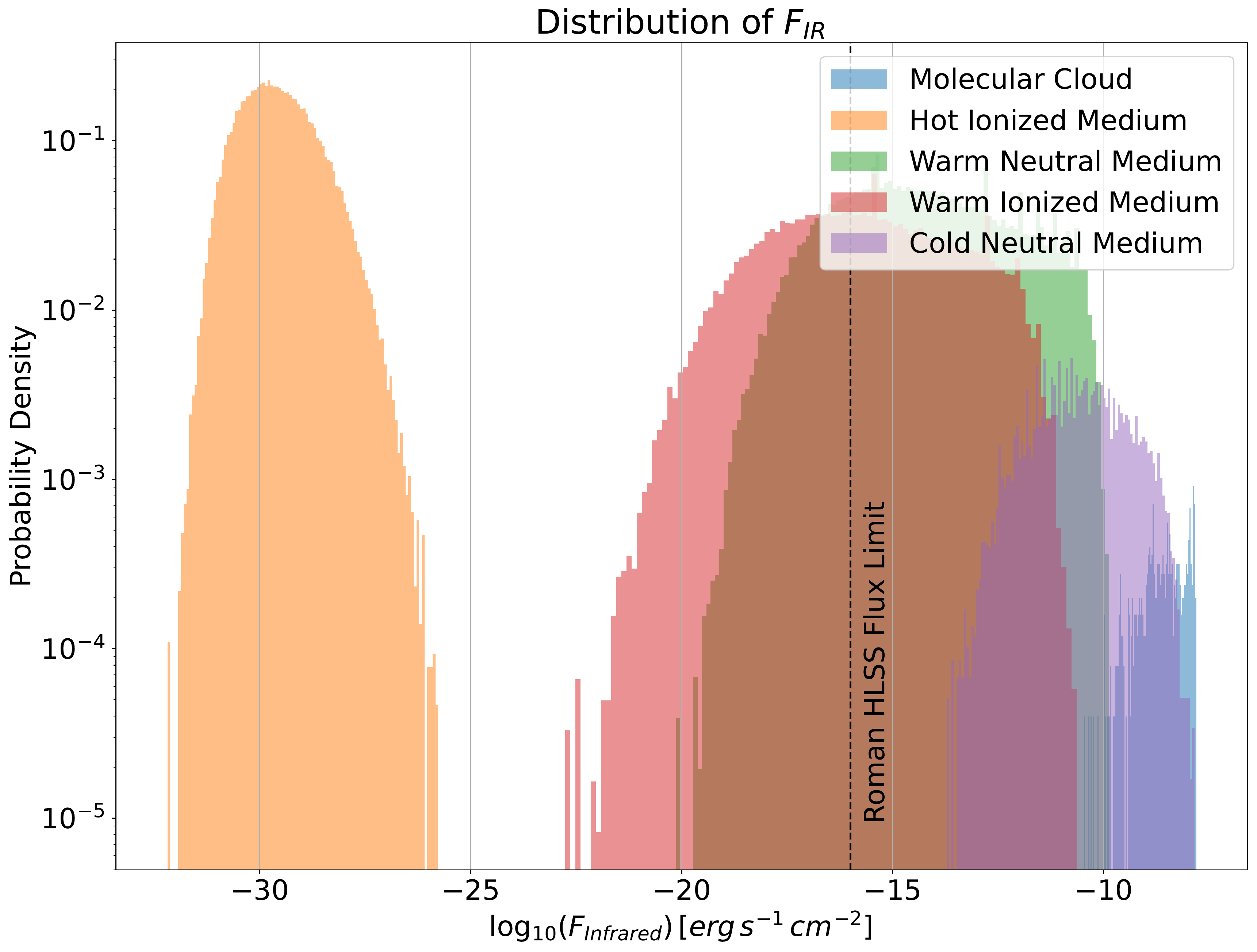}
    \caption{\textbf{Left panel:} X-ray fluxes predicted for IMBHs in the MW, categorized in the five ISM environments shown in the legend. Flux limits for Chandra, e-ROSITA and AXIS are shown. \textbf{Right panel:} Same as the left panel, but for the NIR emission. Flux limit for the High Latitude Spectroscopic Survey (HLSS) of the Roman Space Telescope is shown.}
    \label{fig:XandIR}
\end{figure*}

Figure \ref{fig:XandIR} (left panel) displays the resulting volume-weighted flux distribution in the X-ray band. Our results suggest that a fraction of the population of IMBHs in the MW are detectable by both eROSITA and the Chandra X-ray Observatory in the $0.2-10$ keV band. Specifically, $\sim 27\%$ of IMBHs in the MW could be detected in the X-rays  with Chandra and $\sim 13\%$ with eROSITA. Additionally, AXIS in its proposed intermediate survey \citep{Marchesi_2020} should be able to detect $\sim 38\%$ of wandering IMBHs in the MW. As evident from the categorization in Fig. \ref{fig:XandIR}, fluxes whose emission is generated in MC and CNM environments are observable, as well as a fraction of fluxes generated in WNM and WIM. No fluxes generated within HIM, the most common environment of the ISM, are observable.

Similarly, Fig. \ref{fig:XandIR} (right panel) displays a volume-weighted flux distribution at NIR wavelengths. We found that $\sim 37\%$ of IMBH sources located in the MW could be detected by Roman Space Telescope, with its HLSS survey. The distribution of detectability within the five ISM environments is similar to that described in the X-ray case. 

Overall, X-ray and NIR detectabilities are in the $13\% - 37\%$ range, with very similar distributions among the environments of the ISM.

\subsection{Sub-millimeter and Radio Emission from IMBHs}
\label{subsec:radio}
\begin{figure*}
\includegraphics[width=\columnwidth]{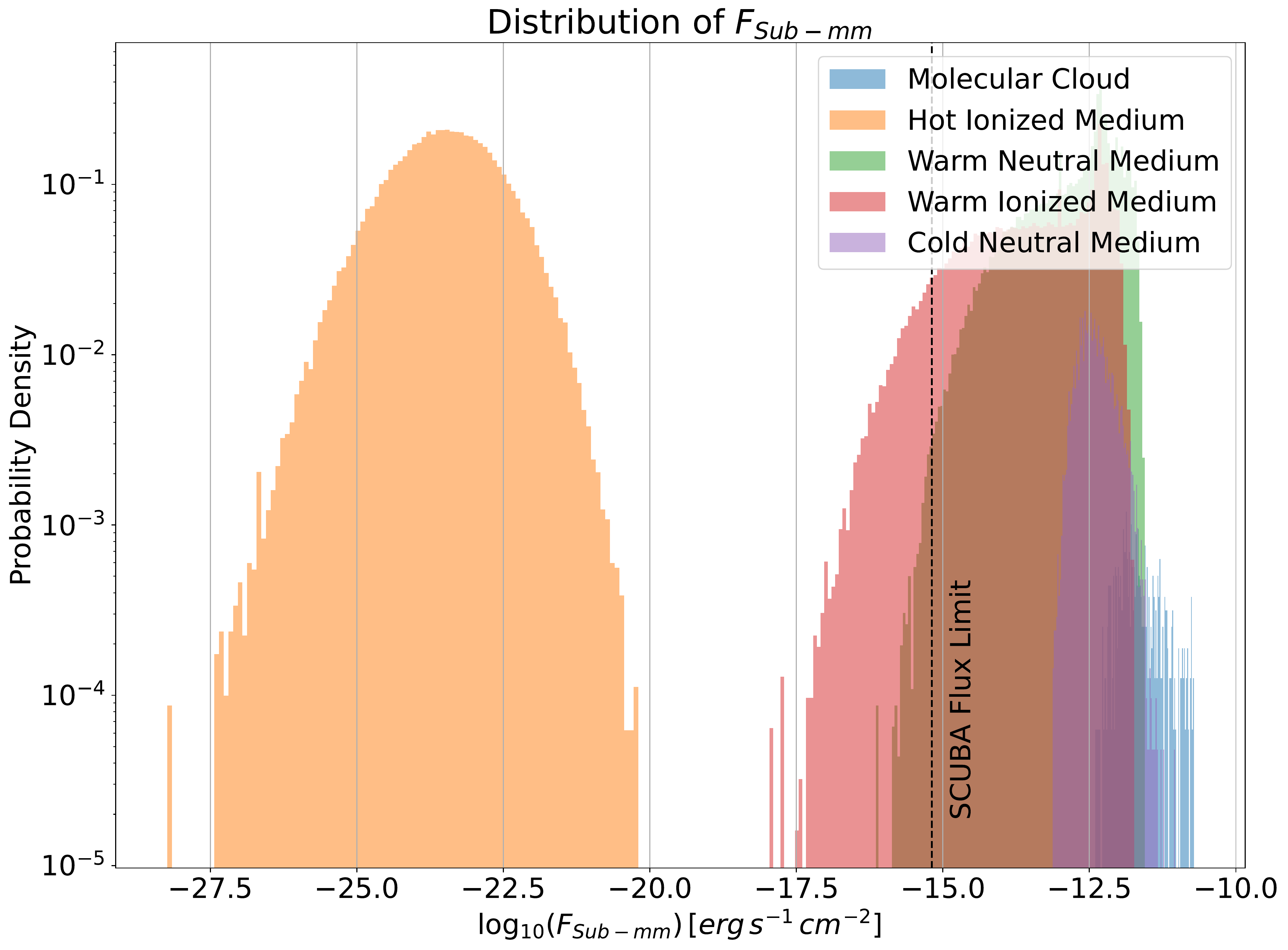}
\includegraphics[width=\columnwidth]{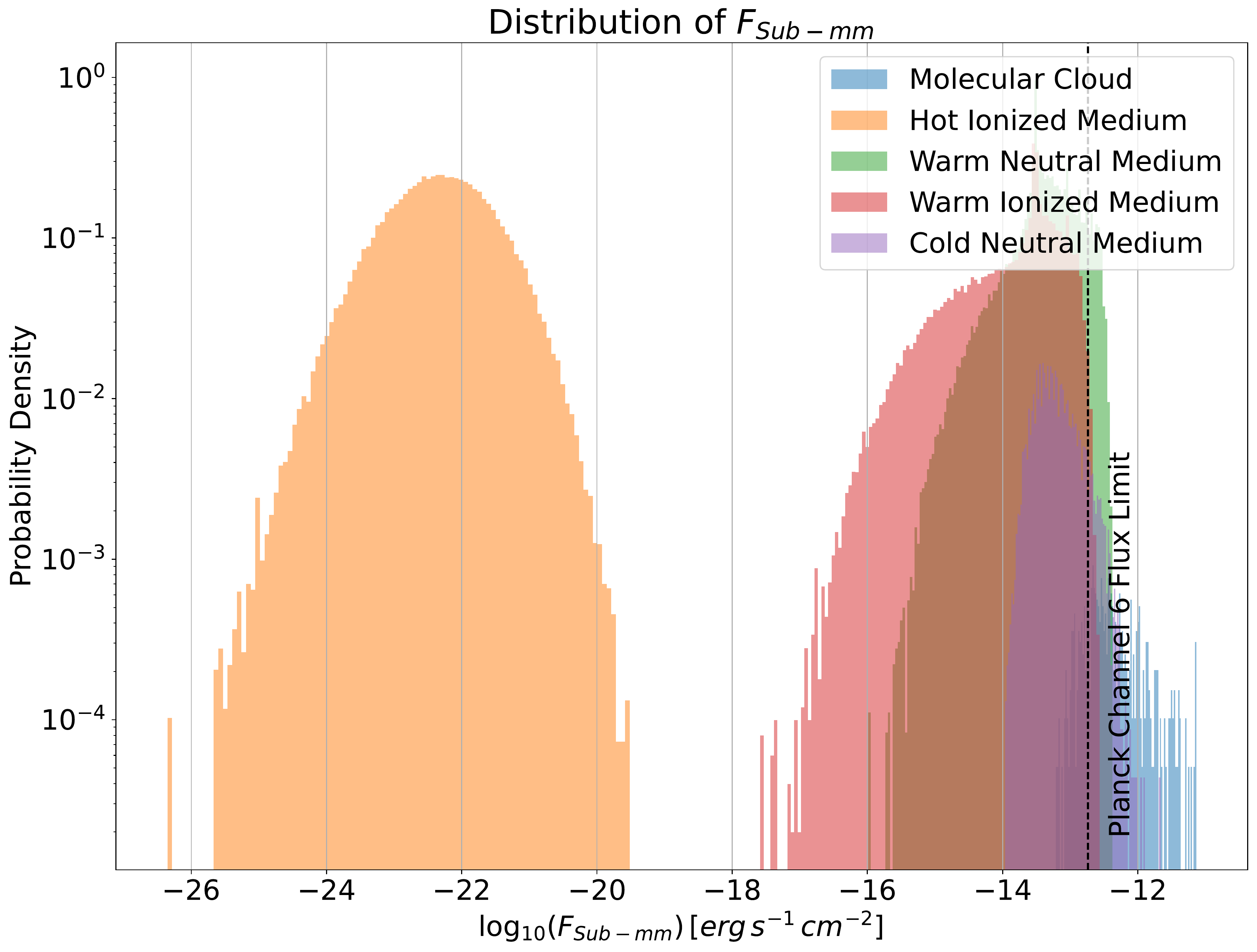}
\includegraphics[width=\columnwidth]{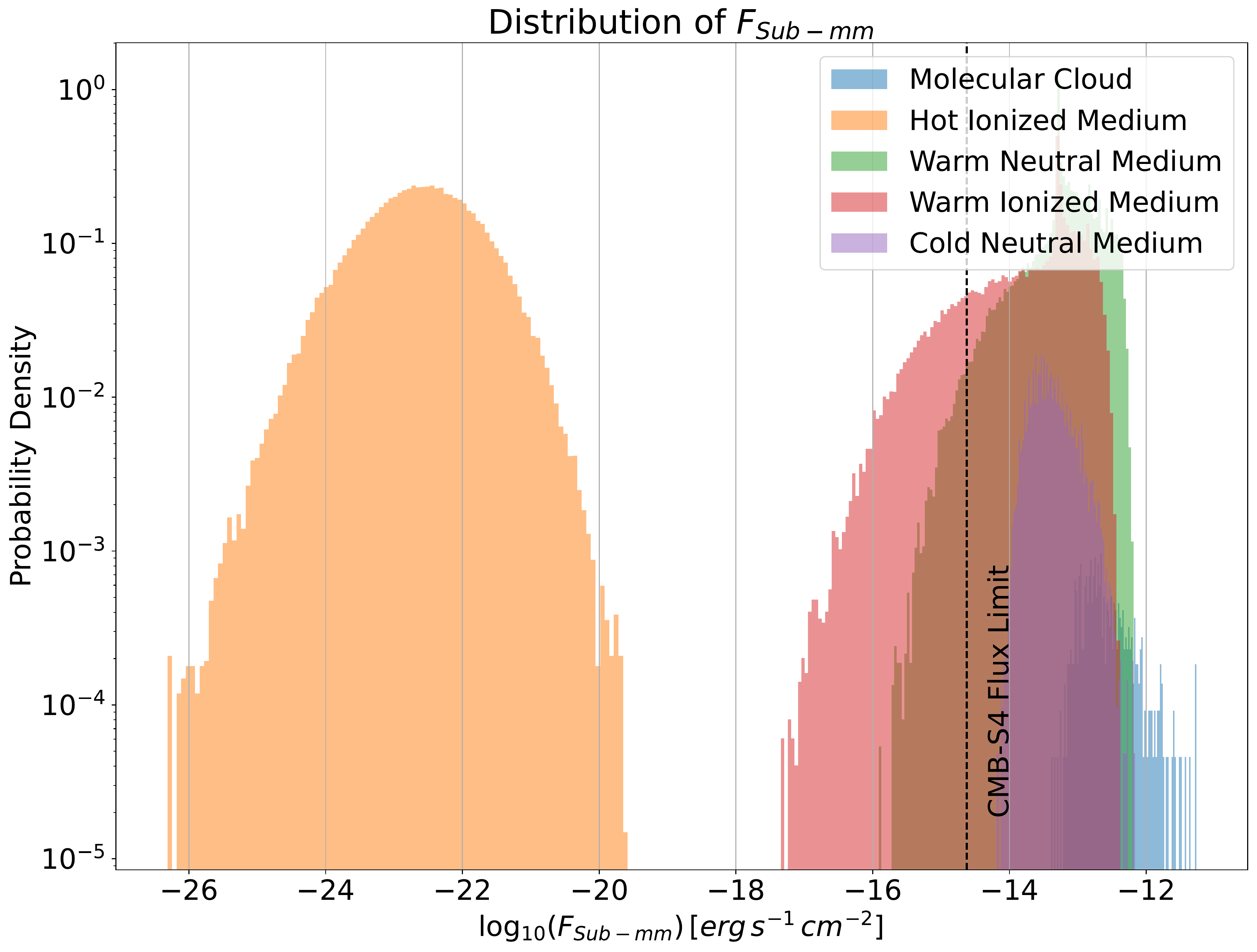}
\includegraphics[width=\columnwidth]{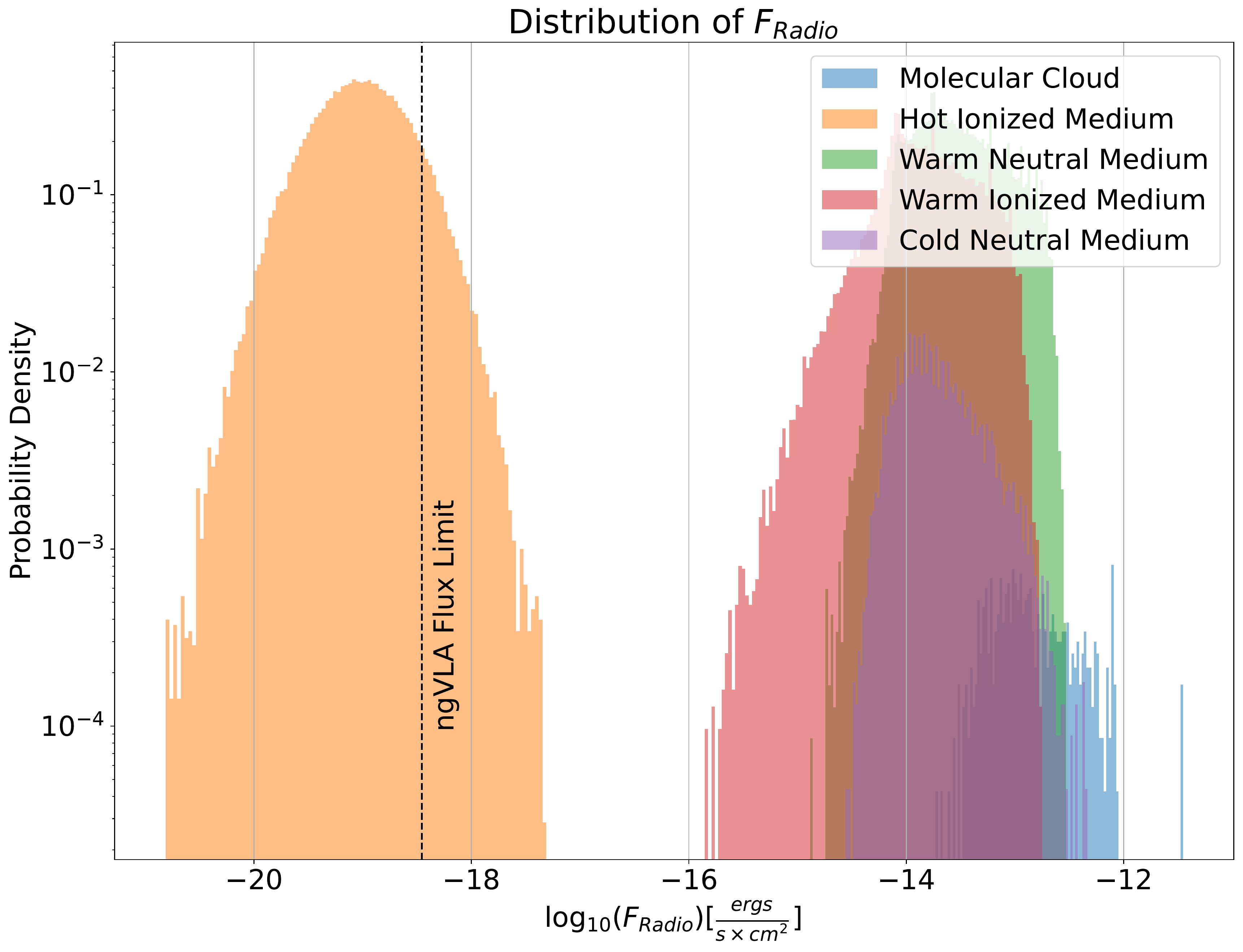}
    \caption{As in Fig. \ref{fig:XandIR}, but for the radio and sub-mm frequency range. \textbf{Top row:} SCUBA (left panel) and Planck Channel 6 (right panel). \textbf{Bottom row:} CMB-S4 (left panel) and ngVLA (right panel). The flux limits of the respective detectors are shown as vertical dashed lines.}
    \label{fig:submm}
\end{figure*}

Figure \ref{fig:submm} displays radio and sub-mm flux distributions for the 4 detectors considered in this section of the EM spectrum. The sub-mm range is shown in different plots, because the frequency ranges covered by SCUBA, Planck and CMB-S4 are slightly different (see Table \ref{table:detectors}). We found that only $\sim 2.9\%$ of IMBHs located in the MW could be detected by Planck, specifically in its channel 6 detector. On the contrary, SCUBA is predicted to detect up to $\sim 51\%$ of the potential IMBH population. Finally, with CMB-S4 in its channel 11 we predict a detectability fraction of $\sim 49\%$. Remarkably, the CMB-S4 will perform a very wide survey, and has a sensitivity sufficient to detect wandering IMBH of $\sim 10^5 \Msun$ in the MC, CNM, WNM and WIM environments.

In the radio, we predict that the ngVLA will be able to detect a remarkable fraction of $\sim 57\%$. The ngVLA is very sensitive, but its field of view is quite narrow, in the range $0.6-25$ arcmin (FWHM) depending on the frequency band \citep{ngVLA}. This, in practice, limits the ngVLA to targeted observations.

Considerations about the overall instrument sensitivity drove us to select ngVLA as the optimal choice for these observations. In fact, with a sensitivity $\sim 10$ times higher than that of the VLA and even of the Atacama Large Millimeter/submillimeter Array (ALMA, see, e.g.,  \citealt{ngVLA}), ngVLA is the only detector able to spot wandering IMBH of $10^5 \Msun$ in the HIM environment, which is the most common one in the MW.

Overall, radio and sub-mm detectors appear to be particularly suited for these observations, with detectability fractions that are in the range $49\% - 57\%$ range (except for Planck, which is lower).

\subsection{Detectability by Instrument and by ISM Environment}

A summary of the detectability fractions for the instruments considered is shown in Table \ref{table:observabilities}, both for $p=0.5$ (our baseline scenario) and for $p=0.75$ (see Eq. \ref{eq:Bondi_corr}).
\begin{table}
    \centering
    \begin{tabular}{c|c|c}
    \hline
\textbf{Detector} & \pbox{20cm}{\textbf{Detectability} \\ \, \, ($p=0.5$)} & \pbox{20cm}{\textbf{Detectability} \\ \, \, ($p=0.75$)} \\
 \hline
         \textbf{Chandra} & $\sim 27\%$ & $\sim 2\%$ \\
         \textbf{eROSITA} & $\sim 13\%$ & $\sim 1\%$ \\
         \textbf{AXIS} & $\sim 38\%$ & $\sim 4\%$ \\
         \textbf{Roman HLSS} & $\sim 37 \%$ & $\sim 2.4\%$ \\
         \textbf{Planck (Channel 6)} & $\sim 2.9\%$ & $\sim 0\%$ \\
         \textbf{SCUBA} & $\sim 51\%$ & $\sim 13\%$ \\
         \textbf{CMB-S4 (Channel 11)} & $\sim 49\%$ & $\sim 5\%$ \\
         \textbf{ngVLA} & $\sim 57\%$ & $\sim 52\%$ \\
\hline
    \end{tabular}
    \caption{Summary of the detectabilities for all the instruments studied in this work, for two indexes $p$ (see Eq. \ref{eq:Bondi_corr}).}
    \label{table:observabilities}
\end{table}
Detectabilities in the baseline scenario range from $\sim 2.9\%$ to $\sim 57\%$ and the best odds are consistently found for lower energy spectral ranges, from the NIR to the radio. This is due to BHs in ADAF mode emitting a significant fraction of their bolometric luminosity in these bands \citep{Pesce_2021}.
Ultimately, also the X-ray range is moderately favorable, with detectability percentages reaching $\sim 27\%$ with the Chandra X-ray observatory in the $0.2-10$ keV range and $\sim 38\%$ with AXIS in the $0.5-2$ keV range.

In the $p=0.75$ scenario, the detectabilities are significantly lower for all instruments, except for the ngVLA, for which they remain slightly above $50\%$. This further consolidates the ngVLA as the preferred instrument to look for this undetected population of sources.

\begin{figure*}
\center
\begin{tabular}{cc}
\subfloat[Detectabilities of Chandra by ISM Environment]{
    \begin{tikzpicture}  
\begin{axis}  
[  
    ybar,  
    enlargelimits=0.15,  
    ylabel={Chandra X-ray Observatory Observability[\%]},  
    xlabel={\ ISM Environment},  
    symbolic x coords={MC, CNM, WNM, WIM, HIM}, % these are the specification of coordinates on the x-axis.  
    xtick=data,  
     nodes near coords, % this command is used to mention the y-axis points on the top of the particular bar.  
    nodes near coords align={vertical},  
    ]  
\addplot coordinates {(MC,100) (CNM,100) (WNM,60.8) (WIM,37.0) (HIM,0) };  
\end{axis}  
\end{tikzpicture}  
}
&
\subfloat[Detectabilities of eROSITA by ISM Environment]{
    \begin{tikzpicture}  
\begin{axis}  
[  
    ybar,  
    enlargelimits=0.15,  
    ylabel={eROSITA 0.2-10 keV Observability[\%]},  
    xlabel={\ ISM Environment},  
    symbolic x coords={MC, CNM, WNM, WIM, HIM}, % these are the specification of coordinates on the x-axis.  
    xtick=data,  
     nodes near coords, % this command is used to mention the y-axis points on the top of the particular bar.  
    nodes near coords align={vertical},  
    ]  
\addplot coordinates {(MC,100) (CNM,99.4) (WNM,30.0) (WIM,13.7) (HIM,0) };  
\end{axis}  
\end{tikzpicture}  
} \\
\subfloat[Detectabilities of Roman by ISM Environment]{
    \begin{tikzpicture}  
\begin{axis}  
[  
    ybar,  
    enlargelimits=0.15,  
    ylabel={Roman HLSS Observability[\%]},  
    xlabel={\ ISM Environment},  
    symbolic x coords={MC, CNM, WNM, WIM, HIM}, % these are the specification of coordinates on the x-axis.  
    xtick=data,  
     nodes near coords, % this command is used to mention the y-axis points on the top of the particular bar.  
    nodes near coords align={vertical},  
    ]  
\addplot coordinates {(MC,100) (CNM,100) (WNM,79.6) (WIM,54.1) (HIM,0) };  
\end{axis}  
\end{tikzpicture}  
}
&
\subfloat[Detectabilities of SCUBA by ISM Environment]{
    \begin{tikzpicture}  
\begin{axis}  
[  
    ybar,  
    enlargelimits=0.15,  
    ylabel={SCUBA Observability[\%]},  
    xlabel={\ ISM Environment},  
    symbolic x coords={MC, CNM, WNM, WIM, HIM}, % these are the specification of coordinates on the x-axis.  
    xtick=data,  
     nodes near coords, % this command is used to mention the y-axis points on the top of the particular bar.  
    nodes near coords align={vertical}, 
    ]  
\addplot coordinates {(MC,100) (CNM,100) (WNM,99.8) (WIM, 92.8) (HIM,0) };  
\end{axis}  
\end{tikzpicture}  
}
\\
\subfloat[Detectabilities of CMB-S4 by ISM Environment]{
    \begin{tikzpicture}  
\begin{axis}  
[  
    ybar,  
    enlargelimits=0.15,  
    ylabel={CMB-S4 Observability[\%]},  
    xlabel={\ ISM Environment},  
    symbolic x coords={MC, CNM, WNM, WIM, HIM}, % these are the specification of coordinates on the x-axis.  
    xtick=data,  
     nodes near coords, % this command is used to mention the y-axis points on the top of the particular bar.  
    nodes near coords align={vertical},  
    ]  
\addplot coordinates {(MC,100) (CNM,100) (WNM,98.3) (WIM,84.1) (HIM,0) };  
  
\end{axis}  
\end{tikzpicture}  
}
&
\subfloat[Detectabilities of ngVLA by ISM Environment]{
    \begin{tikzpicture}  
\begin{axis}  
[  
    ybar,  
    enlargelimits=0.15,  
    ylabel={ngVLA Observability[\%]},  
    xlabel={\ ISM Environment},  
    symbolic x coords={MC, CNM, WNM, WIM, HIM}, % these are the specification of coordinates on the x-axis.  
    xtick=data,  
     nodes near coords, % this command is used to mention the y-axis points on the top of the particular bar.  
    nodes near coords align={vertical}, 
    ]  
\addplot coordinates {(MC, 100) (CNM, 100) (WNM,100) (WIM, 100) (HIM,9.2) };  
\end{axis}  
\end{tikzpicture}  
}
\end{tabular}
\caption{Detectability fraction for some representative detectors, categorized in the five ISM environments.}
\label{fig:ISM_detector}
\end{figure*}
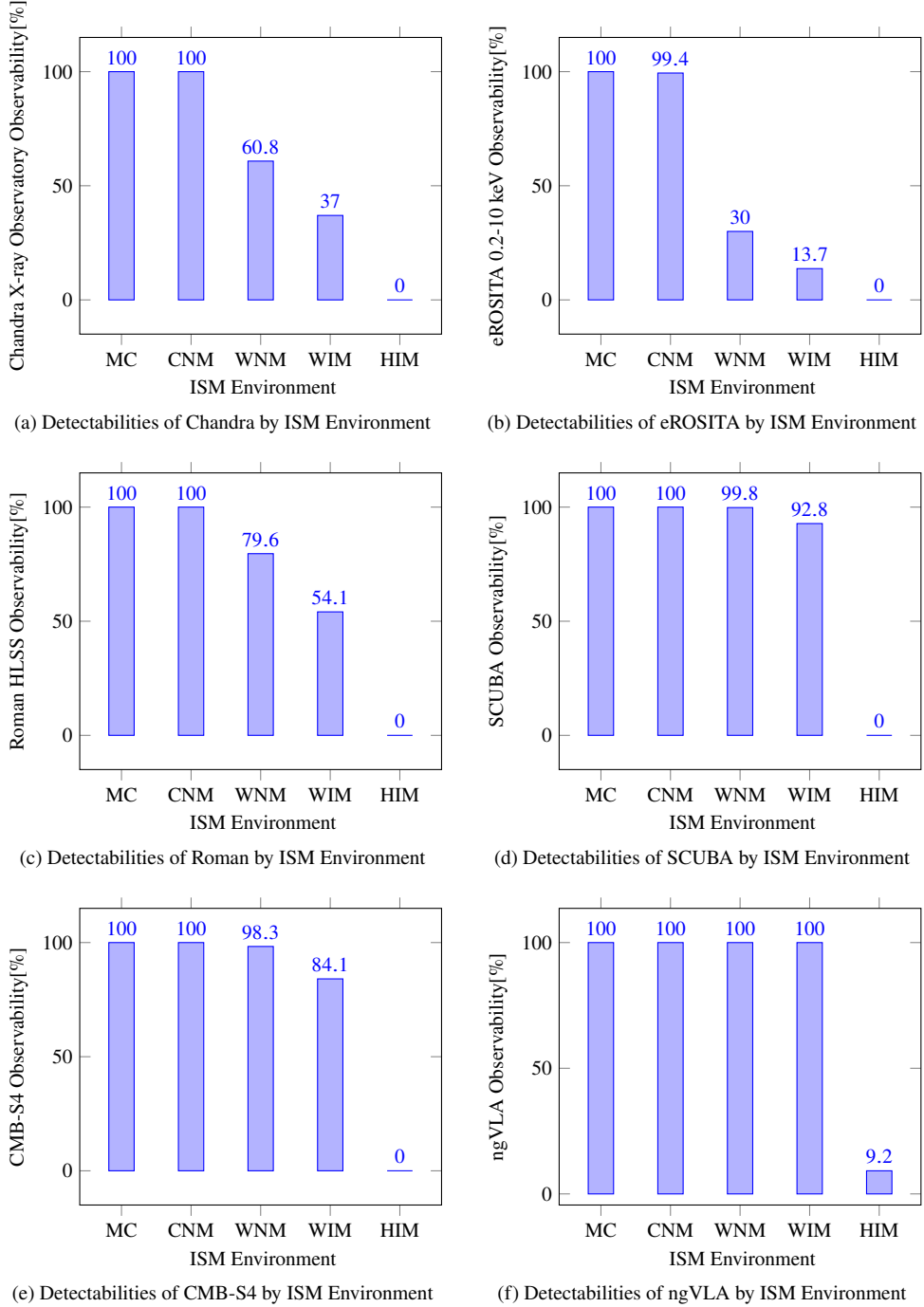

To further study which ISM environments are more conducive to possible detections, in Fig. \ref{fig:ISM_detector} we display the detectabilities for each environment. In this figure we include Chandra, eROSITA, Roman, SCUBA, CMB-S4 and the ngVLA.
Irrespective of the detector, the MC environment shows consistently the best chances for observation, with detection fractions $\sim 100\%$ in every band.
Assuming a typical MC diameter of $\sim 100$ pc, the crossing time of an IMBH in the MW could be of the order of $\sim 1$ Myr. It is possible that IMBHs could be practically observed only during these rare, but very bright, events. Note, however, that these events would not be ``transient'' in the common meaning of the term, given the very long time scales involved, although variations in the local gas density of the MC could produce an observable light curve.
On the other hand, MCs are known to have hydrogen column densities in excess of $\sim 10^{21-22} \, \mathrm{cm^{-2}}$ \citep{Schneider_2015} in the their deepest regions. These typical column densities are lower than the Compton-thick regime ($N_{\rm H} = 1.5\times 10^{24} \, \mathrm{cm^{-2}}$), hence we do not expect a significant absorption of the X-ray photons emitted in the accretion process.
The second best environment to detect IMBHs is, as expected, the CNM with detection rates also $\sim 100\%$.

Overall, Fig. \ref{fig:ISM_detector} confirms that SCUBA and the ngVLA are the best instruments to study wandering IMBHs in the MW, with detectability fractions consistently $\gtrsim 50\%$ in every ISM environment, except the HIM, which leads to detection rates $\sim 0\%$ for all instruments apart from the ngVLA (for which it is $\sim 9\%$).
The ngVLA, as pointed out in \S \ref{subsec:radio}, has a sensitivity $\sim 10$ times higher than ALMA, and is the only detector able to search for wandering IMBHs in the HIM.

SCUBA has been operating for a long time, and has already collected large amounts of data in the sub-millimeter. No clear detection of wandering IMBHs is reported to date. This is either due to the fact that they are absent from the MW, or we simply did not have clear prescriptions to select candidates. We provide some selection criteria, also in the sub-mm band, in the forthcoming \S \ref{subsec:selection}.

\subsection{Selection Criteria for IMBHs}
\label{subsec:selection}
Our study suggests that wandering IMBHs in the MW could be detected in the X-ray, NIR, sub-mm and radio bands, especially when transiting inside MC or CNM environments. 
The fact that they are detectable is just a statement of possibility, but the actual selection of sources in wide-area, multi-wavelength observations is poised to be challenging.

\begin{figure}
\includegraphics[width=\columnwidth]{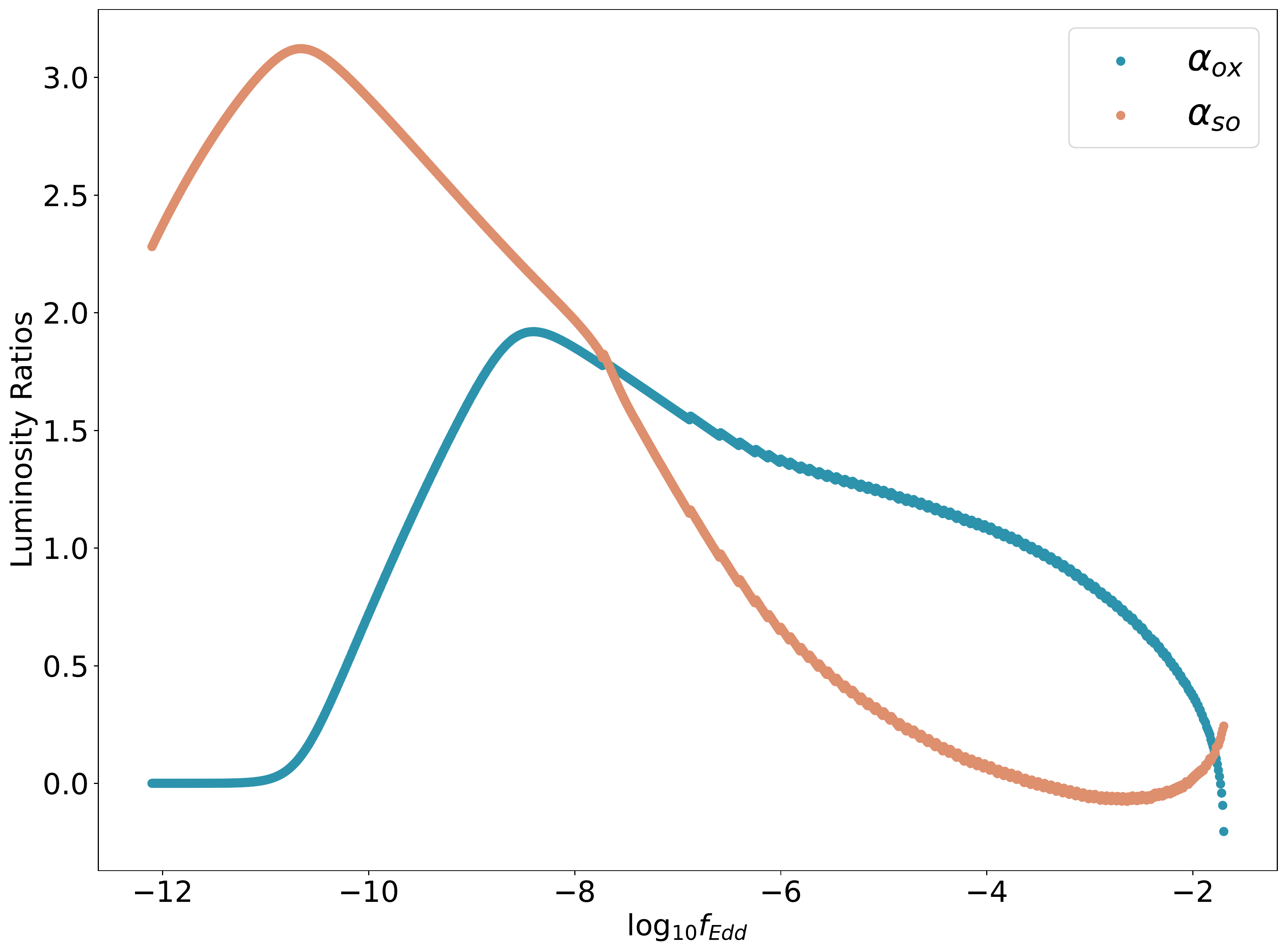}
    \caption{Luminosity ratios (X-ray to optical/UV and optical/UV to sub-mm), as a function of the Eddington ratio.}
    \label{fig:lum_ratios}
\end{figure}

To facilitate this task, we present some basic selection criteria that could be used in photometric surveys of sources in the MW to pre-select objects of interest. In particular, we present two luminosity ratios, which are displayed in Fig. \ref{fig:lum_ratios} as a function of the Eddington ratio $f_{\rm Edd}$ of the IMBH. Of course, the SED produced by an IMBH will change as a function of the accretion rate that characterizes it. Hence, it is not possible to provide a single value of the luminosity ratios that we expect from a wandering $10^5 \Msun$ IMBH, but rather a range of ratios. For a given multi-wavelength photometric observation of a single IMBH candidate, both the black hole mass and the Eddington ratio would be fixed, hence a calculation of both ratios we provide can be instrumental to: (i) assess whether or not the source is a viable IMBH candidate, and (ii) estimate its Eddington ratio.

The two luminosity ratios that we considered are: (i) the X-ray to optical/UV ratio in terms of the standard $\alpha_{\rm ox}$ parameter (see, e.g., \citealt{Lusso_2010}), and (ii) an optical/UV to sub-mm ratio.
The $\alpha_{\rm ox}$ is defined as:
\begin{equation}
    \alpha_{\rm ox} = -\frac{\log_{10}(L_{2 \, \mathrm{keV}} / L_{2500 \, \mathrm{\angstrom}})}{2.605} \, ,
\end{equation}
and typically assumes values in the range $1-2$ for Type 1 AGN \citep{Lusso_2010}.
We define the optical/UV to sub-mm ratio as:
\begin{equation}
    \alpha_{\rm so} = -\frac{\log_{10} (L_{2500 \, \mathrm{\angstrom}}/L_{1 \, \mathrm{mm}})}{3.602} \, .
\end{equation}
The Eddington ratio varies between $10^{-12}$ and $10^{-1.7}$, which is the maximum value allowed in our ADAF framework (see \S \ref{subsec:SED}). 

Figure \ref{fig:lum_ratios} shows a peak of $\alpha_{\rm ox} \sim 1.9$ for Eddington ratios $\sim 10^{-8.4}$, decreasing to $\alpha_{\rm ox} \sim 0.4$ for rates of $\sim 1\%$ of Eddington. For $\alpha_{\rm so}$ we find that the curve peaks at $\sim 3.1$ for Eddington ratios $\sim 10^{-11}$, also decreasing to $\sim 0.4$ for rates of $\sim 1\%$ of Eddington. The latter ratio shows that the sub-mm emission of wandering IMBHs, for $f_{\rm Edd} < 10^{-8}$, is significantly higher than the optical, UV, and X-ray emission.

Also, Fig. \ref{fig:lum_ratios} shows that the Eddington ratio, which varies over 10 orders of magnitude in our analysis, strongly influences the SED shapes (see also Fig. \ref{fig:SED_display}), hence the luminosity ratios.

A Galactic source displaying luminosity ratios significantly different from the ones shown in Fig. \ref{fig:lum_ratios} can be explained by several means. For example, a transient source can be a tidal-distruption event. A non-transient source could be a wandering IMBH with a mass significantly different from $10^5 \Msun$, or an IMBH accreting in non-ADAF mode.

\subsection{Constraints on the Galactic IMBH Population}
\label{subsec:constraints}
The nearly $100\%$ detectabilities of wandering $10^5\Msun$ IMBHs in MC and CNM environments can lead to placing upper limits on the overall population of these objects in the gaseous regions of the MW. 

These upper limits are based on the following assumption: as the transit of a $10^5 \Msun$ wandering IMBH in a MC or CNM environments would produce X-ray fluxes $\sim 5 - 10$ orders of magnitude higher than the Chandra sensitivity, it is reasonable to conjecture that such objects have not crossed these two regions, at least in the recent past. Hence, the reciprocal of their respective volume fractions in the MW provides an estimate of their maximum total number, assuming a uniform distribution of a population of $N_\bullet$ IMBHs of $10^5 \Msun$ in the MW.
\begin{itemize}
    \item From the volume fraction of the MC environment, ($\sim 0.05\%$), we obtain an upper limit of $N_\bullet < 2000$. 
    \item From the volume fraction of the CNM environment, ($\sim 1.0\%$), we obtain an upper limit of $N_\bullet < 100$. 
\end{itemize}
Of course, these upper limits are tentative, as it is possible that such IMBHs may have transited in MC and CNM environments in the past, when observing facilities were not available. In fact, as we have previously noted, transit events inside MC environments would typically last $\sim 1$ Myr.

The predicted accretion rates of IMBHs in MC environments are very large, with X-ray luminosities in excess of $10^{41} \, \mathrm{erg \, s^{-1}}$ for the brightest events. This radiation could significantly impact the stability of the MC. In fact, radiation pressure and photo-ablation effects could, in principle, destroy the MC after a bright period much shorter than the estimated crossing time (see, e.g, \citealt{Goicoechea_2016, Ferrara_2016}).
For X-ray luminosities $\ll 10^{41} \, \mathrm{erg \, s^{-1}}$, possibly generated by lighter IMBHs or in lower density MCs, this emission could be contaminated by star-forming regions. The study of X-ray emission by massive BHs and star-forming regions in galaxies, especially dwarfs, is not new (see, e.g., \citealt{Aird_2017}), and methods have been recently developed to untangle these emission components (see, e.g., \citealt{Mezcua_2018_b, Birchall_2020}) and to estimate the fraction of dwarfs containing a massive BH \citep{Pacucci_2021_active}.

Additional feedback effects on the ISM, such as the creation of jets and lobes around the accreting black hole, should not be excluded. The investigation of the potential effects of IMBHs on the ISM will be investigated in future work.

\section{Discussion and Conclusions}
\label{sec:disc}
IMBHs are the only population of BHs that have not been detected in our Galaxy. 
As most of the volume in the MW lacks a sufficient reservoir of gas to trigger large accretion rates, it is unsurprising that these objects, if they exist, are undetected to date.

The presence of a wandering IMBH in the MW Galaxy would cause several effects, some of which may be detectable. First, a stellar density cusp, with typical radial density profile $\rho \propto r^{-7/4}$, is expected to form around massive black holes \citep{Bahcall_Wolf_1976, Greene_2021_wandering}.
Although this enhanced stellar density is generally challenging to detect directly, it can bear consequences which are observable, e.g., a different spatial distributions of neutron stars compared to a model without an IMBH \citep{Loeb_IMBH_Tuc_2017}.

Additionally, the passage of these sources along the line of sight with a background star will create a micro-lensing event, which could be detected with the use of extensive stellar surveys, such as Gaia (see, e.g., \citealt{Toki_2021}). Other techniques, such as the symmetric achromatic variability \citep{Vedantham_2017}, could be used to detect IMBHs via milli-lensing.

An additional detection venue, the one we investigate in this paper, requires that the IMBH accretes some gas from the ISM, likely in ADAF mode, and thus radiates at low levels. This study showed, using five realistic environments in the MW and a tailored code to compute the SED of black holes accreting in ADAF mode, that this detection is possible, over a very wide range of frequencies in the EM spectrum, from the X-rays to radio.

Our two main results are the following:
\begin{itemize}
    \item IMBHs are more likely to be detected in infrared, sub-mm and radio bands, with the Roman telescope (NIR) able to detect $\sim 37\%$ of the population, the future CMB-S4 (sub-mm) detecting $\sim 49\%$ of the population, the SCUBA telescope (sub-mm) detecting $\sim 51\%$, and the ngVLA (radio) detecting $\sim 57\%$. The latter detectability fraction holds even with the higher value of the index $p=0.75$.
    \item The two MW environments more likely to lead to a detection are molecular clouds and cold neutral medium. Both, accounting for only $\sim 0.05\%$ and $\sim 1\%$ of the MW volume, respectively, lead to detection in the vast majority ($>99\%)$ of cases, in all the frequency ranges studied.
\end{itemize}

It is interesting to note that the large fluxes predicted for transits in the very rare CNM and MC environments ($\sim 5$ and $\sim 10$ orders of magnitude higher than the Chandra flux limit, respectively) pose significant constraints on detection scenarios. In particular, IMBHs in these uncommon MW environments should be characterized by relatively short bursts of detectability, which may have hindered their observation thus far. Additionally, the fact that the densest environments in the MW are preferentially located towards the Galactic center favors a detection strategy based on wide-field surveys directed towards the central $\sim 1$ kpc region of the MW. Remarkably, this is also the region where simulations show that the highest density of wandering IMBHs reside \citep{Weller_2022}.

For example, a Galactic plane survey with the ngVLA, using the $93 \, \mathrm{GHz}$ band, with a maximum angular resolution of $\sim 10^{-4}$ arcsec, would have a sensitivity of $0.4 \, \mathrm{\mu Jy/beam}$. This flux limit would be achieved with 1 hour integration time, making an extended survey of the Galactic plane possible. Such a survey would be able to detect $\sim 50\%$ of the putative $10^5 \Msun$ population of IMBH in the Galactic plane.
 
Regarding other, more common environments such as the WNM, the WIM and the HIM, the lack of a clear detection thus far can be more likely attributable to the need of specific selection criteria to distinguish them from more common sources. The selection criteria presented in this paper provide useful foundations to look for these less striking sources.

In summary, this study provided a blueprint for the detection of IMBHs in the MW, due to the radiation from ISM accretion. Of course, the actual detection of these sources remains challenging, but, we showed, plausible in most of the EM spectrum.
The hunt for the only remaining undetected BH population in the MW is open: our study can hopefully guide the search with current and future observational facilities.

\section*{Acknowledgements}
We thank the anonymous referee for constructive comments on the manuscript.
B.S.S. acknowledges undergraduate research support provided in part by the Harvard College Program for Research in Science and Engineering (PRISE).
F.P. acknowledges support from a Clay Fellowship administered by the Smithsonian Astrophysical Observatory. 
R.N. acknowledges support from NSF grants OISE-1743747 and AST1816420.
The Authors thank Dom Pesce for providing the code to calculate the SEDs.
This work was partly performed at the Aspen Center for Physics, which is supported by National Science Foundation grant PHY-1607611. The participation of F.P. at the Aspen Center for Physics was supported by the Simons Foundation.
This work was also supported by the Black Hole Initiative at Harvard University, which is funded by grants from the John Templeton Foundation and the Gordon and Betty Moore Foundation.

%%%%%%%%%%%%%%%%%%%%%%%%%%%%%%%%%%%%%%%%%%%%%%%%%%
\section*{Data Availability}

The code used to analyze the data will be shared on reasonable request to the corresponding authors.



\bibliographystyle{mnras}
\bibliography{ms} % if your bibtex file is called example.bib

%\begin{thebibliography}{99}
%\end{thebibliography}

%%%%%%%%%%%%%%%%%%%%%%%%%%%%%%%%%%%%%%%%%%%%%%%%%%

%%%%%%%%%%%%%%%%% APPENDICES %%%%%%%%%%%%%%%%%%%%%

%%%%%%%%%%%%%%%%%%%%%%%%%%%%%%%%%%%%%%%%%%%%%%%%%%

% Don't change these lines
\bsp	% typesetting comment
\label{lastpage}
\end{document}